\documentclass[preprint,sort&compress,3p,11pt,times]{elsarticle}

\makeatletter
\def\ps@pprintTitle{%
 \let\@oddhead\@empty
 \let\@evenhead\@empty
 \def\@oddfoot{\emph{Published by Computer Physics Communications, https://doi.org/10.1016/j.cpc.2024.109327}\hfill}%
 \let\@evenfoot\@oddfoot}

\usepackage{amssymb}
\usepackage{amsmath}
\usepackage{url}
\usepackage{hyperref}
\usepackage{xcolor}
\usepackage{algorithm}
\usepackage{algpseudocode}
\usepackage{algorithmicx}
\usepackage{graphicx}
\usepackage{subfig}
\usepackage{lineno}

\usepackage{listings}
\newcounter{bla}

\journal{Computer Physics Communications}

\begin{document}

\begin{frontmatter}

\title{libEMMI\_MGFD: A program of marine controlled-source electromagnetic modelling and inversion using frequency-domain multigrid solver}

\author[address1]{Pengliang Yang\corref{correspondence}}
\ead{ypl.2100@gmail.com}
\author[address1]{An Ping}
\ead{2021110084@stu.hit.edu.cn}

\cortext[correspondence]{Corresponding author.}
\address[address1]{School of Mathematics,  Harbin Institute of Technology,  150001,  China}

\begin{abstract}
  We develop a  software package  libEMMI\_MGFD for 3D frequency-domain marine controlled-source electromagnetic (CSEM) modelling and inversion. It is the first open-source C program tailored for geometrical multigrid (GMG) CSEM simulation. An volumetric anisotropic averaging scheme has been employed to compute effective medium for modelling over uniform and nonuniform grid. The computing coordinate is aligned with acquisition geometry by rotation with the azimuth and dip angles, facilitating the injection of the source and the extraction of data with arbitrary orientations. Efficient nonlinear optimization is achieved using quasi-Newton scheme assisted with bisection backtracking line search. In constructing the modularized Maxwell solver and evaluating the misfit and gradient for 3D CSEM inversion, the reverse communication technique is the key to the compaction of the software while maintaining the computational performance. A number of numeric tests demonstrate the efficiency of the modelling while preserving the solution accuracy. A 3D marine CSEM inversion example has been examined for resistivity imaging.
\end{abstract}


\end{frontmatter}


{\bf PROGRAM SUMMARY}

\begin{small}
\noindent
{\em Program Title:} libEMMI\_MGFD                                \\
{\em CPC Library link to program files:} (to be added by Technical Editor) \\
{\em Developer's repository link:} \url{https://github.com/yangpl/libEMMI_MGFD} \\
{\em Code Ocean capsule:} (to be added by Technical Editor)\\
{\em Licensing provisions:} GNU General Public License v3.0 \\
{\em Programming language:} C, Shell  \\
{\em Operating System:} Linux \\
{\em Software dependencies:} MPI \cite{1}\\
{\em Nature of problem:}
3D Controlled-source electromagnetic modelling and inversion\\
{\em Solution method:}
Frequency-domain modelling on staggered grid;
Geometrical multigrid (GMG) method as an iterative solver;
Quasi-Newton method for nonlinear minimization;
Bisection-based backtracking line search\\

\end{small}

\section{Introduction}

Controlled-source electromagnetic (CSEM) modelling and inversion have emerged as invaluable tools in understanding subsurface geological structures and resource exploration. This technique utilizes controlled electromagnetic sources, typically deployed on the seafloor or land surface, to probe the Earth's subsurface. By analyzing the response of the Earth's electrical conductivity to these electromagnetic signals, CSEM enables the detection and characterization of various subsurface features, including hydrocarbon reservoirs, mineral deposits, and geological formations \cite{constable2006mapping,constable2007introduction}.

A variety of algorithms and computing techniques have been routinely employed for electromagnetic modelling, e.g., finite difference method in frequency domain
\citep{newman1995frequency,smith1996conservative1,mulder2006multigrid,streich20093d} and in time domain \citep{oristaglio1984diffusion,wang1993finite,Taflove_2005_CEF,Mittet_2010_HFD}, as well as
finite element method in frequency domain\citep{li20072d,da2012finite,key2016mare2dem,rochlitz2019custem}. The discretization of the frequency-domain Maxwell equation in 3D leads to a linear system possessing extremely large number of unknowns. Solution of such a linear system is computationally intensive and storage demanding. The use of direct solver to solve the diffusive Maxwell equation is straightforward \cite{streich20093d,grayver2013gji}, but may require significant amount of computer memory, which is rather challenging if the size of the problem becomes prohibitively large.
The iterative methods such as QMR \cite{newman1995frequency,cai20143d}, BICGSTAB \cite{smith1996conservative2}, GMRES \cite{puzyrev2013parallel} and multigrid \citep{mulder2006multigrid,haber2007octree,koldan2014algebraic} algorithms are commonplaces to alleviate memory issues, but may suffer from convergence difficulties due to the ill-conditioning of the linear system  \cite{Yang_2023_HFDNU,Yang_2023_3dcsem}.

CSEM inversion, sometimes referred to as resistivity imaging, plays a crucial role in CSEM data interpretation by reconstructing subsurface conductivity/resistivity distributions from measured electromagnetic field data. Using efficient modelling engines, inversion algorithms seek to minimize the difference between observed and predicted electromagnetic data by iteratively adjusting model parameters, such as subsurface conductivity values. This iterative optimization process aims to find the most plausible subsurface conductivity distribution that best fits the observed data, thereby providing valuable insights into subsurface structures and properties.

There have been a large number of open software for electromagnetic modelling and inversion, e.g.  \verb|SimPEG| \cite{cockett2015simpeg}, \verb|MARE2DEM| \cite{key2016mare2dem},  \verb|pyGIMLI| \cite{rucker2017pygimli},  \verb|emg3d| \cite{werthmuller2019emg3d},
,  \verb|custEM| \cite{rochlitz2019custem}, \verb|ResIPy| \cite{blanchy2020resipy}, just to name a few. Among them, some of them are dedicated to marine CSEM, for example, \verb|MARE2DEM| and \verb|emg3d|.  We aim to make some relevant contributions in this aspect, by developing a standalone  software \verb|libEMMI_MGFD| for 3D large scale problems in high performance parallel computing settings.
\begin{itemize}
  \item
  Different from our previous code \verb|libEMMI| \cite{Yang_2023_3dcsem} computing frequency domain responses via time-domain modelling, the core modelling engine of this package solves diffusive Maxwell equation discretized in the frequency domain.
  Inspired by the optimal convergence within $O(N)$ complexity, we re-implemented the geometrical multigrid (GMG) modelling of 3D CSEM data proposed by \cite{mulder2006multigrid} in C programming language, following the \verb|emg3d| code in Python \cite{werthmuller2019emg3d}.

  \item
  The goal of CSEM inversion is to retrieve resistivity anomaly starting from a crude initial guess  by data driven approaches. Due to the high dimensionality of the parameter space, it is infeasible to use global optimization methods to fit such a parametric model.   Local optimization with quasi-Newton and Newton type method naturally becomes the method of choice.

\item
  The software has been carefully designed thanks to the reverse communication strategy. It provides a succinct yet reliable mechanism to modularize the GMG linear solver and the code blocks for function and gradient evaluation which are frequently used in every iteration of the nonlinear inversion. Based on such a concept, the LBFGS algorithm combined with bisection-based backtracking line search are coded compactly.

\end{itemize}
Despite the fact that modelling EM diffusion using GMG by \cite{mulder2006multigrid} and the LBFGS-based CSEM inversion  \cite{Plessix_2008_RIC} are known, we highlight that binding these components into to construct a complete set of parallel and efficient open source software in C is still worthwhile for a large number of computational geophysicists to do further development.

\section{Methodology}

\subsection{CSEM forward problem}

The physics of controlled-source electromagnetic (CSEM) technology is governed by diffusive Maxwell equation, consisting of Faraday's law and Amp\`ere's law. The two laws combined with proper boundary condition forms a set of partial differential equation (PDE) in the frequency domain
\begin{subequations}\label{eq:maxwelleh}
  \begin{align}
    \nabla\times E  -\mathrm{i}\omega \mu H &= M_s,\label{eq:faraday}\\
    \nabla\times H  -\sigma E &= J_s, \label{eq:ampere}
    \end{align}
\end{subequations}
where the electrical field $E=(E_x, E_y, E_z)^\mathrm{T}$ and the magnetic field $H=(H_x, H_y, H_z)^\mathrm{T}$  are  excited by source current $J_s=(J_x, J_y, J_z)^\mathrm{T}$ and $M_s=(M_x, M_y, M_z)^\mathrm{T}$ based on the electric conductivity $\sigma$ and magnetic permeability $\mu$. 
In practical applications, we consider vanishing magnetic source and constant magnetic permeability, that is, $M_s=0$ and $\mu=\mu_r\mu_0=4\pi\times 10^{-7}$ H/m (with $\mu_r=1$).  In the above, Fourier transform convention $\partial_t \leftrightarrow -\mathrm{i}\omega$ with $\omega$ being the angular frequency has been adopted.

By eliminating the magnetic field $H$, the Maxwell equation translates into
\begin{equation}\label{eq:curlcurl}
  \nabla \times \mu_r^{-1}\nabla \times E  -\mathrm{i}\omega\mu_0\sigma E =\mathrm{i}\omega\mu_0 J_s + \nabla\times\mu_r^{-1} M_s.
\end{equation}
Equation \eqref{eq:curlcurl} can be viewed as a complex-valued linear system
\begin{equation}\label{eq:axb}
AE=b,
\end{equation}
where $A=\nabla \times \mu_r^{-1}\nabla \times  -\mathrm{i}\omega\mu_0\sigma$, $b=\mathrm{i}\omega\mu_0 J_s + \nabla\times\mu_r^{-1} M_s$. The electrical field $E$ is obtained from the solution of the above linear system using direct or iterative solvers. Away from sources, the magnetic field $H$ can then be computed from equation \eqref{eq:faraday}
\begin{equation}\label{eq:fromEtoH}
H = (\mathrm{i}\omega\mu)^{-1}\nabla\times E.
\end{equation}

\subsection{Frequency-domain geometrical multigrid modelling}

\begin{figure}
  \centering
  \includegraphics[width=0.8\linewidth]{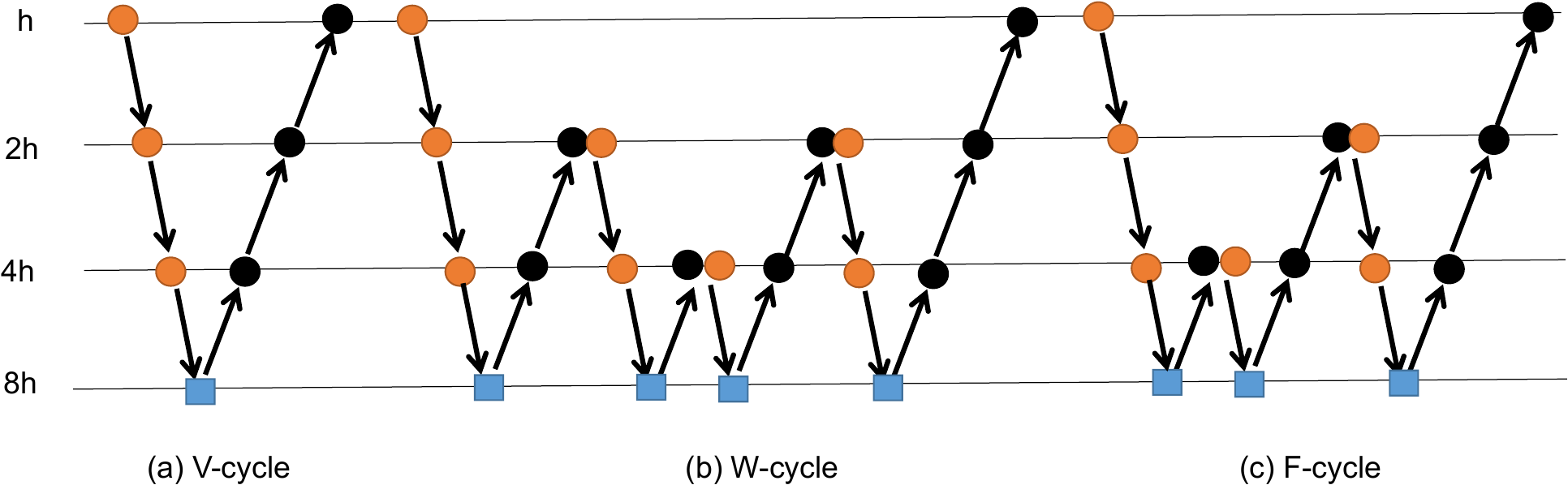}
  \caption{Schematic illustration of (a) V cycle, (b) W cycle and (c) F cycle based on 4 grid levels. The $j$th grid level corresponds to a discretized cell size  $2^jh$ ($j=0$ corresponds to the finest grid). The arrow downwards $\searrow$ indicates a restriction from the fine grid $2^{j-1}h$ to the fine grid $2^jh$, while the arrow upwards $\nearrow$ indicates a prolongation from the fine grid $2^jh$ to the coarse grid $2^{j-1}h$. The orange and black dots stand for pre- and post- smoothing operations. The blue square at the coarsest grid corresponds to direct solve, which is often done by several times of smoothing.}\label{fig:multigrid}
\end{figure}

To make the meshing simple, equation \eqref{eq:curlcurl} has been discretized on a nonuniform cartesian grid. We choose the geometrical multigrid (GMG) method for the numerical modelling of diffusive Maxwell equation in the frequency domain, inspired by its optimal convergence properties \cite{Briggs_2000_Multigrid}. The core idea of GMG is simple:
\begin{itemize}
\item The high frequency components of the residual of the PDE can be effectively removed using few number of smoothing steps (called relaxation), e.g., Gauss-Seidel algorithm;
\item The low frequency components of the residual of the PDE can then be transformed to higher frequency components by mapping them onto a coarse grid (it is coined restriction);
\item The solution at the coarse grid must be mapped back to fine grid as a correction to the approximate solution at the fine grid in a bottom-to-top fashion (it is coined prolongation or interpolation), to obtain a better approximation of the true solution of the equation;
\item By cascading these steps in a multiscale manner, all frequency components of the residual are expected to be reduced effectively, leading to a significantly refined solution to the PDE.
\end{itemize}
A full chain of the recursion involving multiple grid levels is called V cycle. To further improve the efficiency of error smoothing, W and F cycles can then be constructed using the basic V cycles. The residual of the equation vanishes after a number of V, W or F cycles. Figure~\ref{fig:multigrid} schematically illustrates the computing steps of V, W and F cycles based on recursion among 4 grid levels.

Since $\nabla\times E=\nabla\times (E+\nabla V)$ is valid for any scalar potential field $V$, the solution of the above linear system is therefore non-unique due to the null space of the curl-curl operator, in case the second term on the left hand side of equation \eqref{eq:curlcurl} is extremely small. This is exactly the case when the frequency is very low ($\omega\approx 0$) and the conductivity is negligible ($\sigma \approx 0$, it is typically the case in the air). This leads to a highly ill-conditioned linear system, making the iterative methods difficult to converge. Preconditioning is therefore designed to rectify the poor convergence, using divergence correction \cite{smith1996conservative2} or Helmholtz decomposition (also known as A-$\phi$ potential field decomposition) with Coulomb \cite{haber2000fast} or Lorentz gauge \cite{weiss2013AphiD}. We apply the preconditioning in the process of Gauss-Seidel smoothing, which is the key to the success of GMG in solving diffusive Maxwell equation. A pointwise Gauss-Seidel sweeping in a lexicographic order has been applied following the proposal of \cite{mulder2006multigrid}. At each point, a $6\times 6$ linear system is solved using LU decomposition. The updated electrical field is then injected as the right hand side for the solution of the next $6\times 6$ linear system. Indeed, this block Gauss-Seidel iteration is nothing else than the multiplicative Schwarz preconditioning \cite[chapter 14.3]{Saad_2003_IMS}. Recall that the key idea of preconditioning is to approximately solve the linear system cheaply as effective as possible, the Gauss-Seidel smoothing overcomes the ill-conditioning of the linear system in \eqref{eq:axb}.

The use of nonuniform grid creates certain amount of numerical anisotropy which makes multigrid iterations converge slowly. A line Gauss-Seidel scheme combined with semi-coarsening (halving the number of gridpoints in only two directions in 3D when mapping the residual from fine to coarse grid) is a useful recipe to deal with this problem, by combining all $6\times 6$ small linear systems to construct a larger one along $x$, $y$ and $z$ axes. From a domain decomposition point of view, each linear system along a line forms a subdomain to partition the full 3D grid. Since the number of unknowns are much larger, we resort to incomplete LU (ILU) (instead of Cholesky factorization as done by \cite{mulder2006multigrid}) to solve it: the intermediate elements created during ILU are stored in place thanks to the compressed diagonal storage (CDS) technique \cite{Barret_1994_TSL}. Compared with the $6\times 6$ block, the size of the linear system along a line is much closer to the full 3D grid. Consequently, it is expected to supply a better approximation of the full linear system, leading to faster convergence rate.  We refer to \cite[chapter 3.1.2]{jonsthovel2006improvement} to check out more details on building up the linear system for block and line Gauss-Seidel smoothing.

\subsection{Regridding over the effective medium}

The accuracy of the modelling engine is generally higher on a dense mesh than that on a coarse mesh.
This however significantly increases the computational cost, since the number of grid points to discretize
the same physical domain grows quickly with refined grid spacing. Nonuniform grid has
been widely adopted to reduce the computational cost. To achieve efficient CSEM modelling, the grid spacing in regions demanding high modelling accuracy is made to be small, while the grid spacing in other regions is taken to be large. \verb|libEMMI_MGFD| adopts the power-law grid stretching \cite[Appendix C]{mulder2006multigrid}: the grid size gradually increases away from the source following a geometric progression.  To simulate EM fields propagating to very far distance using limited mesh size, we extend the domain of interest six skin depth, such that the amplitude goes further down to $e^{-6}=0.25\%$ to avoid boundary effect. The generation of the nonuniform grid after domain extension follows the fixed point iteration algorithm in \cite{Yang_2023_HFDNU}.

 To map the input parameters to modelling grid, we perform volume average for the horizontal conductivities $\sigma_{xx}$, $\sigma_{yy}$ and vertical resistivity $\sigma_{zz}^{-1}$ for multigrid modelling. The averaged values at the cell center $(x_{i+0.5},y_{j+0.5},z_{k+0.5})$ are computed 
\begin{subequations}
  \begin{align}
    \bar{\sigma}_{xx}(x_{i+0.5},y_{j+0.5},z_{k+0.5})=&\frac{1}{\bar{V}_{i+0.5,j+0.5,k+0.5}} \int_{x_i}^{x_{i+1}} \int_{y_j}^{y_{j+1}} \int_{z_j}^{z_{j+1}}\sigma_{xx}(x,y,z)\mathrm{d}x\mathrm{d}y\mathrm{d}z,\\
    \bar{\sigma}_{yy}(x_{i+0.5},y_{j+0.5},z_{k+0.5})=&\frac{1}{\bar{V}_{i+0.5,j+0.5,k+0.5}}  \int_{x_j}^{x_{j+1}}\int_{y_j}^{y_{j+1}} \int_{z_j}^{z_{j+1}}\sigma_{yy}(x,y,z)\mathrm{d}x\mathrm{d}y\mathrm{d}z,\\
    \bar{\sigma}_{zz}(x_{i+0.5},y_{j+0.5},z_{k+0.5})=&\left(\frac{1}{\bar{V}_{i+0.5,j+0.5,k+0.5}}  \int_{x_j}^{x_{j+1}} \int_{y_j}^{y_{j+1}}\int_{z_k}^{z_{k+1}}\sigma_{zz}^{-1}(x,y,z)\mathrm{d}x\mathrm{d}y\mathrm{d}z\right)^{-1}
\end{align}
\end{subequations}
over the cell volume $\bar{V}_{i+0.5,j+0.5,k+0.5}=(x_{i+1}-x_i)(y_{i+1}-y_i)(z_{i+1}-z_i)$.
It can be implemented by tensorial product of averaging in each direction. A simple and efficient way of the above averaging for every grid cell is to difference the integral starting from a reference origin $(x_0,y_0,z_0)$. In 1D case, it corresponds to $ \int_{x_i}^{x_{i+1}}= \int_{x_0}^{x_{i+1}}- \int_{x_0}^{x_{i}}$. The above VTI anisotropic averaging relies on the fact that the tangential ﬁelds see the resistors connected in parallel, while the normal ﬁeld sees the resistors connected in series \cite{Yang_2023_libEMM}.

\subsection{CSEM inverse problem}

The CSEM inversion performs least-squares minimization between the observed EM data $d=(d_E,d_H)^\mathrm{T}$ and the synthetic data extracted from the simulated EM field $u=(E,H)^\mathrm{T}$. The total objective function incorporating the data misfit and a model regularization term is specified by
\begin{equation}
 f(m) = \underbrace{\frac{1}{2}\|W_d(d - Ru)\|^2}_{ f_d(m)} + \underbrace{\frac{\beta}{2}\|W_m(m-m_{ref})\|^2}_{ f_m(m)},
\end{equation}
where the inversion parameter $m$ is a function of electrical conductivity $\sigma$; $R$ is the restriction operator mapping the EM field from the full domain to recording locations prescribed by the receivers. The penalty parameter $\beta$ balances the data misfit term $ f_d(m)$ and the model regularization term $ f_m(m)$; $W_d$ and $W_m$ are weighting matrices related to the data uncertainty and the model roughness. To minimize the objective function, we often consider a quadratic approximation of the original misfit function perturbed by a model increment $\delta m$
\begin{equation}
   f(m+\delta m) \approx  f(m) + \frac{\partial  f(m)}{\partial m}\cdot\delta m
  + \frac{1}{2}\delta m\cdot \frac{\partial^2  f(m)}{\partial m^2}\cdot \delta m
\end{equation}
To find the optimal $\delta m$ allowing the misfit attaining its minimum, we requires
$\partial  f(m+\delta m)/\partial \delta m=0$, yielding the amount of model perturbation
\begin{equation}\label{eq:normal}
\delta m=-\left(\frac{\partial^2  f(m)}{\partial m^2}\right)^{-1}\frac{\partial  f(m)}{\partial m},
\end{equation}
where  $\partial^2  f(m)/\partial m^2$ is the Hessian matrix. Assuming a symmetric positive definite (SPD) Hessian, equation \eqref{eq:normal} can be solved using conjugate gradient (CG) method, leading to the so-called Newton-CG method which is computationally expensive \cite{Yang_2018_TRN}.
The quasi-Newton LBFGS algorithm \cite{Nocedal_2006_NOO} approximates the inverse Hessian based on gradient of the misfit $\partial  f(m)/\partial m$. It consists of the first derivatives of $ f_d(m)$ and $ f_m(m)$   with respect to $m$
\begin{subequations}
  \begin{align}
    \frac{\partial  f_d(m)}{\partial m}=&\Re\langle W_d R \frac{\partial {u}}{\partial m}, W_d(Ru-d)\rangle, \label{eq:graddat}\\
    \frac{\partial  f_m(m)}{\partial m} = &\beta W_m^\mathrm{T} W_m (m-m_{ref}),
  \end{align}
\end{subequations}
where $\Re$ takes the real part of a complex value.
Our CSEM inversion takes the logarithm of the resistivity ($\rho:=1/\sigma$ is the inverse of conductivity) as the inversion parameter, that is, $m=\ln\rho$. Here, the gradient of the data misfit in equation \eqref{eq:graddat} with respect to conductivity $\sigma$ can be computed using the adjoint state method \citep{Plessix_2006_RAS,Yang_2023_3dcsem}
\begin{equation}\label{eq:grad}
\frac{\partial  f_d(m)}{\partial \sigma}=-\Re \sum_\omega \underline{\bar{E}}^\mathrm{T}\cdot E,
\end{equation}
where  the overbar takes the complex conjugate of a complex number; $\underline{\bar{E}}$ is the so-called adjoint electrical field satisfying the following equation
\begin{equation}\label{eq:adjeq}
\nabla \times \mu_r^{-1}\nabla \times \underline{\bar{E}} - \mathrm{i}\omega \mu_0\sigma\underline{\bar{E}} = \mathrm{i}\omega \mu_0 \overline{\delta d_E} + \nabla \times \mu_r^{-1} \overline{\delta d_H},
\end{equation}
where $\delta d_E=R^\mathrm{T} W_{d_E}^\mathrm{T}W_{d_E}(d_E-RE)$ and $\delta d_H=R^\mathrm{T}W_{d_H}^\mathrm{T}W_{d_H} (d_H-RH)$ are the electrical and magnetic components of the data residuals weighted by $W_d=\mbox{diag}(W_{d_E},W_{d_H})$. The adjoint equation \eqref{eq:adjeq} is of exactly the same form as the forward equation \eqref{eq:curlcurl} except that the conjugation of the data residuals $\overline{\delta d_E}$ and $\overline{\delta d_H}$ replaces the role of $J_s$ and $M_s$, implying that the same modelling engine can be used to compute the adjoint field $\underline{\bar{E}}$.

Since the CSEM data varies several order of magnitude within several kilometers offset, the normalization of the CSEM data plays a crucial role for the success of the inversion, based on the measurement uncertainty of the acquired electrical and magnetic fields.
 Taking into account the imperfection of the instruments and the ambient noise in the real acquisition, a practical value is around 3\%, while the noise floor for electrical field and magnetic field are approximately $n_E=10^{-15}$ V/m and $n_H=10^{-13}$ Tesla \cite{mittet2012detection}.  The data weighting matrix $W_d$ is then taken to be a diagonal matrix, with the diagonal elements being the inverse of the uncertainty.

\subsection{Coordinate transformation}

\begin{figure}
  \centering
  \includegraphics[width=0.7\linewidth]{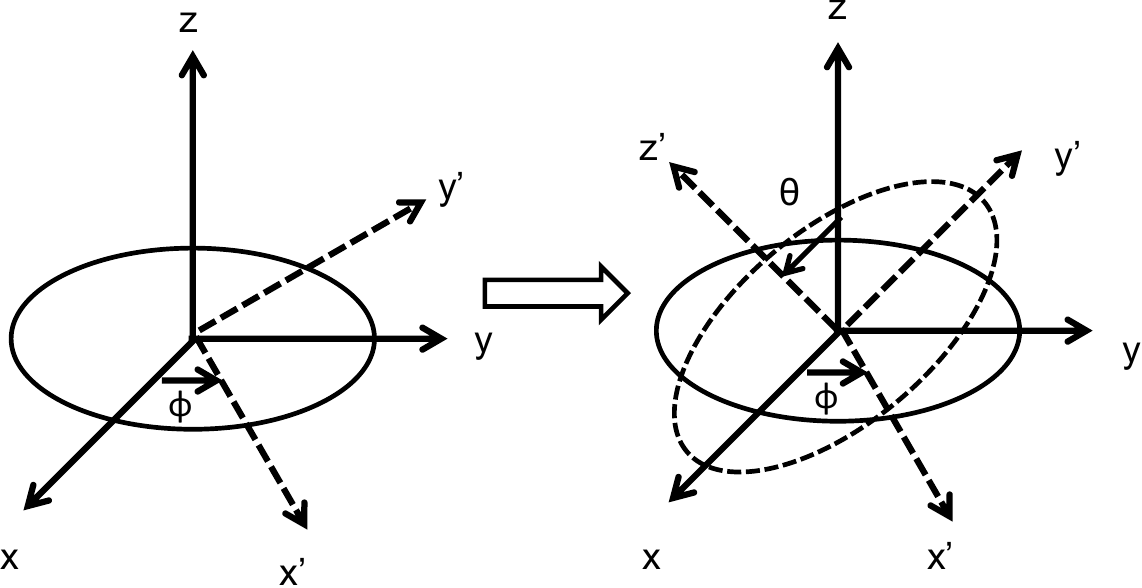}
  \caption{The EM data can be aligned with the computing coordinate $(x,y,z)$ by rotation of the data in the acquisition coordinate $(x',y',z')$ along $x-y$ plane up to an azimuth angle $\phi$ first, and then along $z$ axis with a dip angle $\theta$.}\label{fig:rotation}
\end{figure}

The computing coordinate used for numerical modelling and the original coordinate used for data acquisition are normally not the same. As shown in Figure~\ref{fig:rotation}, the EM data in  the computing coordinate $(x,y,z)$ can be aligned with  the acquisition coordinate $(x',y',z')$ by first rotating along $x-y$ plane up to an azimuth angle $\phi$, and then along $z$ axis with a dip angle $\theta$. As a result, the field $F=E,H$ in the two coordinate systems can be related to each other through
\begin{equation}\label{eq:rotation}
\begin{bmatrix}
F_{x'} \\
F_{y'} \\
F_{z'}
\end{bmatrix}=C\begin{bmatrix}
F_x \\
F_y \\
F_z
\end{bmatrix},
\end{equation}
where the rotation matrix is specified by 
\begin{equation}
  C=\begin{bmatrix}
1 & 0 & 0 \\
0  & \cos \theta  &  \sin \theta \\
0 &  -\sin \theta &  \cos \theta
\end{bmatrix}\begin{bmatrix}
\cos \phi  &  \sin \phi  & 0 \\
-\sin \phi &  \cos \phi  & 0 \\
0 & 0 & 1
\end{bmatrix}
=\begin{bmatrix}
\cos \phi            &  \sin\phi             & 0 \\
-\sin\phi \cos\theta &  \cos\phi \cos\theta  & \sin\theta \\
\sin\phi  \sin\theta & -\cos\phi \sin\theta  & \cos\theta
\end{bmatrix}.
\end{equation}
Equation \eqref{eq:rotation} supplies a recipe to extract the modelling data in the computing grid to correctly match the observation coordinate.

Conversely, one can convert the fields from the acquisition coordinate $(x',y',z')$ to  the computing coordinate $(x,y,z)$ via
\begin{equation}\label{eq:Rinv}
\begin{bmatrix}
F_x \\
F_y \\
F_z
\end{bmatrix}=C^{-1} \begin{bmatrix}
F_{x'} \\
F_{y'} \\
F_{z'}
\end{bmatrix},
\end{equation}
where
\begin{equation}
C^{-1}=C^T=\begin{bmatrix}
\cos \phi  &  -\sin\phi \cos\theta & \sin\phi \sin\theta \\
\sin\phi   &  \cos\phi \cos\theta  & -\cos\phi\sin\theta \\
0          &           \sin\theta  & \cos\theta
\end{bmatrix}.
\end{equation}
Equation \eqref{eq:Rinv} allows us to inject the EM sources in the original acquisition coordinate into the computing grid.

\section{Software implementation}

The \verb|libEMMI_MGFD| code inherits a significant number of features from our fictitious wave domain CSEM imaging software \verb|libEMMI| \cite{Yang_2023_libEMM,Yang_2023_3dcsem} and the seismic full waveform inversion (FWI) software \verb|SMIwiz| \cite{Yang_2024_SMIwiz}. The pointers associated with specific data structures are widely used in \verb|libEMMI_MGFD|, in order to parse the parameters succinctly and pass them into dedicated subroutines for computation.
A pointer of type \verb|acq_t| holds the information related to the data acquisition geometry. A pointer of type \verb|emf_t| indexes all information of electromagnetic field for forward and adjoint modelling. A pointer of type \verb|fwi_t| is designed for inversion purposes (here we interpret EM inversion as full waveform inversion based on the electromagnetic equation). A pointer of type \verb|opt_t| is dedicated to do nonlinear optimization.
The interested reader are referred to \verb|libEMMI| and \verb|SMIwiz| for a detailed description of the common features aforementioned.

In the following, we focus on a modularized programming of an iterative solution of frequency-domain Maxwell equation using multigrid method in \verb|libEMMI_MGFD|, which clearly differs from \verb|libEMMI| modelling based on the fictitious wave domain time stepping scheme \cite{Yang_2023_HFDNU}. We code quasi-Newton LBFGS algorithm using bisection line search in the C programming language. The key to achieving an efficient and successful implementation is the use of reverse communication.

\subsection{Recursive V, W and F cycles}

We define a data of type \verb|gmg_t| to encapsulate the grid and medium properties of 3D EM fields used in geometrical multigrid modelling:
\begin{lstlisting}
typedef struct{
  int n1, n2, n3;//number of cells along x, y and z 
  double *x1, *x2, *x3;//nodes along x, y and z
  double *x1s, *x2s, *x3s;//staggered nodes along x, y and z
  double *d1, *d2, *d3;//cell sizes centered at integer nodes
  double *d1s, *d2s, *d3s;//cell sizes centered at staggered nodes
  complex ****u, ****f, ****r;//r=f-Au, u=(Ex,Ey,Ez)^T
  double ***sigma11, ***sigma22, ***sigma33;//electric conductivities
  double ***invmur;//inverse of relative magnetic permeability
} gmg_t;
\end{lstlisting}
The dimensionalities \verb|n1|, \verb|n2| and \verb|n3| must be multiple of a factor of power of 2, in that the size of grid along $x$, $y$ and $z$ directions must be regularly halved  from fine to coarse grid each time, according to the basic principle of GMG method. The maximum number of grid levels is  \verb|lmax=min(m1,m2,m3)| with \verb|m1|, \verb|m2| and \verb|m3| being the largest power of 2 as an integer factor of \verb|n1|, \verb|n2| and \verb|n3|. The electrical conductivities \verb|sigma11|, \verb|sigma22| and \verb|sigma33| and the inverse of relative magnetic permeability \verb|invmur| are 3D arrays of size \verb|n1*n2*n3| defined at cell centers. Since a number of frequencies may be simulated over 3D staggered grid, the solution of the Maxwell equation, its right hand side and the residual of PDE at each grid level are complex-valued 4D arrays stored in \verb|gmg[lev].u|, \verb|gmg[lev].f| and \verb|gmg[lev].r|, where the level index \verb|lev| ranges from 0 to \verb|lmax-1|.

The V cycle is the basic building block for multigrid iterative solver. At each level, we first perform \verb|v1| times Gauss-Seidel smoothing, then map the residual of the Maxwell equation from current level to a coarser grid using \verb|residual(...)| and \verb|restriction(...)|. After calling \verb|v_cycle(...)| recursively, the coarse grid solution  supplies a correction term to update the solution at current level thanks to the routine \verb|prolongation(...)|. A post smoothing of \verb|v2| times allows V cycle moving to an upper level. This idea has been illustrated in the following.

\begin{lstlisting}
void v_cycle(gmg_t *gmg, int lev)
{

  int n, i;

  if(cycleopt==1 && lev==0){//compute the norm of the residual
    residual(gmg, lev);//residual r=f-Au at lev-th lev
    n = 3*(gmg[lev].n1+1)*(gmg[lev].n2+1)*(gmg[lev].n3+1);
    rnorm = sqrt(creal(inner_product(n, &gmg[lev].r[0][0][0][0], &gmg[lev].r[0][0][0][0])));
    if(verb) printf("icycle=%d rnorm=%e\n", icycle, rnorm);

    if(icycle==0) rnorm0 = rnorm;
    else if(rnorm<rnorm0*tol) { icycle=ncycle; return; }
  }

  for(i=0; i<v1; i++) smoothing(gmg, lev, i);//pre-smoothing of u based on u,f at lev-th level
  if(lev<lmax-1 && gmg[lev+1].sc[0]*gmg[lev+1].sc[1]*gmg[lev+1].sc[2]>1){
    residual(gmg, lev);//residual r=f-Au at lev-th lev
    restriction(gmg, lev);//restrict gmg[lev].r to gmg[lev+1].f 

    n = 3*(gmg[lev+1].n1+1)*(gmg[lev+1].n2+1)*(gmg[lev+1].n3+1);
    memset(&gmg[lev+1].u[0][0][0][0], 0, n*sizeof(complex));
    v_cycle(gmg, lev+1);// another v-cycle at (lev+1)-th level

    prolongation(gmg, lev);//interpolate r^h=gmg[lev+1].u to r^2h from (lev+1) to lev-th level
  }
  //if lev==lmax-1, then nx=ny=2, grid size=3*3, only 1 point at the center is unknwn
  //direct solve (equivalent to smoothing at center point) by one post-smoothing will do the job  
  for(i=0; i<v2; i++) smoothing(gmg, lev, i);//post-smoothing
}
\end{lstlisting}
At the outset of V cycle, the norm of the equation residual $r=b-AE$ has been computed using an inner product between complex-valued vectors.
The W and F cycles can then be constructed using V cycle, see \verb|w_cycle(...)| and \verb|f_cycle(...)| for implementation details. The F cycle is the default option due to the efficiency of error smoothing, while W cycle is not recommended because of dramatically increased computational cost. In practice, the iterations over V/F cycles will be terminated if the residual is reduced down 6 orders of magnitude.

\subsection{Modularized multigrid solver}\label{sec:multigrid}

All the related parameters for multigrid modelling of a particular frequency are initialized by the routine \verb|gmg_init(...)|.
The actual modelling will be carried out by calling the routine \verb|gmg_apply(...)|. Note that the input arguments for \verb|gmg_apply(...)| is rather terse, which may use some  parameters from the a copy of the pointer \verb|emf| declared internally thanks to the initialization done by \verb|gmg_init(...)| at the outset of GMG modelling. This allows us to strictly follow the same argument list every time when a GMG must be performed.  Following the same concept, we initialize the pointer \verb|emf| with \verb|emf_init(...)| before multigrid modelling and deallocate the memory  with \verb|emf_close(...)| afterwards. Looping over a number of frequencies based on the modularized multigrid solver leads to a rather compact code for CSEM modelling:
\begin{lstlisting}
  for(ifreq=0; ifreq<emf->nfreq; ifreq++){
    emf_init(acq, emf, ifreq);  
    gmg_init(emf, ifreq);

    n = 3*emf->n123pad;
    x = alloc1complex(n);//vector E=(Ex,Ey,Ez)^T
    b = alloc1complex(n);//vector Js=(Jx,Jy,Jz)^T
    inject_source(acq, emf, b, ifreq);//initialize b=i*omega*mu*Js
    memset(x, 0, n*sizeof(complex));//initialize x=0

    gmg_apply(n, b, x);//after multigrid convergence, x=(Ex,Ey,Ez)^T
    
    extract_emf_data(acq, emf, x, b, ifreq);
    extract_emf_field(acq, emf, x, b, ifreq);

    free1complex(x);
    free1complex(b);
    
    gmg_close();
    emf_close(emf);
  }
\end{lstlisting}

The modularized programming hides the tedious details of  V/F cycles involving Gauss-Seidel smoothing, restriction, prolongation etc. Because of the finite integration formulation, we multiply the cell volume on both sides of the equation. The solution after a number of V/F cycles will be copied into the unknown vector, as demonstrated by the following code snippet.
\begin{lstlisting}
/*< apply multigrid as a linear solver for Ax=b >*/
void gmg_apply(int n, complex *b, complex *x)
{
  int i, j, k;
  int icycle, lev;
  double vol;

  memcpy(&gmg[0].f[0][0][0][0], b, n*sizeof(complex));
  memset(&gmg[0].u[0][0][0][0], 0, n*sizeof(complex));
  for(k=0; k<gmg[0].n3; k++){
    for(j=0; j<gmg[0].n2; j++){
      for(i=0; i<gmg[0].n1; i++){
	//multiply volume 
	vol = gmg[0].d1s[i]*gmg[0].d2s[j]*gmg[0].d3s[k];
	gmg[0].sigma11[k][j][i] *= vol;
	gmg[0].sigma22[k][j][i] *= vol;
	gmg[0].sigma33[k][j][i] *= vol;
	gmg[0].invmur[k][j][i] *= vol;
      }
    }
  }
  for(k=0; k<=gmg[0].n3; k++){
    for(j=0; j<=gmg[0].n2; j++){
      for(i=0; i<=gmg[0].n1; i++){
	//multiply volume on both left and right sides
	vol = gmg[0].d1s[i]*gmg[0].d2[j]*gmg[0].d3[k];
	gmg[0].f[0][k][j][i] *= vol;
	vol = gmg[0].d1[i]*gmg[0].d2s[j]*gmg[0].d3[k];
	gmg[0].f[1][k][j][i] *= vol;
	vol = gmg[0].d1[i]*gmg[0].d2[j]*gmg[0].d3s[k];
	gmg[0].f[2][k][j][i] *= vol;
      }
    }
  }

  for(icycle=0; icycle<ncycle; icycle++){
    for(lev=1; lev<lmax; lev++) grid_init(gmg, lev);
    if(cycleopt==1) v_cycle(gmg, 0);
    if(cycleopt==2) f_cycle(gmg, 0);
    if(cycleopt==3) w_cycle(gmg, 0);
    for(lev=1; lev<lmax; lev++) grid_close(gmg, lev);
  }
  memcpy(x, &gmg[0].u[0][0][0][0], n*sizeof(complex));//copy E into x
  compute_H_from_E(gmg, 0);//compute H and store it in gmg[0].f
  memcpy(b, &gmg[0].f[0][0][0][0], n*sizeof(complex));//copy H into b
}
\end{lstlisting}
In the above code segment, \verb|isemicoarsen| takes the option 1 (to do semi-coarsening) or 0 (for a full coarsening scheme); \verb|tol| (with the default value $10^{-6}$) is the tolerance criterion for V, W and F cycles. At the end of the code, the electric fields $(E_x,E_y,E_z)$ are copied into the vector \verb|x| while the magnetic field $(H_x,H_y,H_z)$ (deduced from the electrical fields according to equation \eqref{eq:fromEtoH}) is stored in the vector \verb|b|.

\subsection{Building the inversion gradient}

The electrical fields at the nodes of the uniform grid are required to construct the inversion gradient according to \eqref{eq:grad}. They can be extracted using trilinear interpolation. Note that the parallel components of the electrical field $E_x$, $E_y$, vertical electrical current $J_z$ and the magnetic components $H_x$, $H_y$ and $H_z$ are continuous in the presence of conductivity discontinuities. Since interpolation assumes the continuity of a given function, we interpolate over $J_z$ and then convert $J_z$ to $E_z$ component by dividing the conductivities for the nodal values. These are embedded in the routine \verb|extract_emf_data(...)| and \verb|extract_emf_field(...)|.

The extracted forward and adjoint EM fields on uniform grid for \verb|nfreq| frequencies will be stored in 4D arrays \verb|E1f,E2f,E3f| and \verb|E1a,E2a,E3a|. Consider VTI anisotropy, we end up with \verb|fwi->npar=2| parameters (horizontal resistivity $\rho_{11}=\rho_{22}$ and vertical resistivity $\rho_{33}$)  in total.
Because the model above the sea floor has been filled with sea water whose conductivity/resistivity is known, there is no need to update the parameters for this part throughout the whole inversion. We therefore mute gradient above the seabed, whose depth is specified by the bathymetry \verb|emf->bathy|. Note that the resistivity parameters are defined at the cell center. In the vicinity of the source within the radius \verb|emf->dsmute|, it is important to mute the gradient to prevent the side effect of singular values affecting the model update. Finally, an MPI reduction sums over the local gradient computed from different sources to form the whole gradient of the data misfit.

\subsection{Quasi-Newton algorithm}

With the gradient at hand,  CSEM inversion can estimate an update vector $\delta m$ according to \eqref{eq:normal}. The limited-memory BFGS (LBFGS) algorithm approximates the inverse Hessian with a matrix $B_k$ up to a scaling factor $\alpha_k$, so that the minimization updates the inversion parameters
\begin{equation}
  m^{k+1}=m^k-\alpha_k B_k \nabla  f^k,
\end{equation}
where $\nabla f^k:=\partial  f(m)/\partial m|_{m=m^k}$ is the gradient of the misfit at the $k$th iteration. Define
\begin{equation}
  \begin{cases}
  s_k :=& m^{k+1}-m^{k}, \\
  y_k :=& \nabla  f^{k+1}- \nabla  f^k,\\
  \rho_k:=&1/(y_k^\mathrm{T} s_k),\\
  V_k :=& I-\rho_k y_k s_k^\mathrm{T}.
  \end{cases}
\end{equation}
The matrix $B_k$ is constructed using the gradients computed in the previous $\ell$ iterations \cite{Nocedal_2006_NOO}
\begin{equation}
  \begin{split}
    B_k = &\gamma_k (V_{k-1}^\mathrm{T} \cdots V_{k-\ell}^\mathrm{T}) (V_{k-\ell}\cdots V_{k-1})\\
    & +\rho_{k-\ell} (V_{k-1}^\mathrm{T} \cdots V_{k-\ell+1}^\mathrm{T}) s_{k-\ell} s_{k-\ell}^\mathrm{T} (V_{k-\ell+1}\cdots V_{k-1}) \\
    &+ \cdots + \rho_{k-1} s_{k-1}s_{k-1}^\mathrm{T},
    \end{split}
\end{equation}
where $\gamma_k=s_{k-1}^\mathrm{T}y_{k-1}/y_{k-1}^\mathrm{T}y_{k-1}$.
Instead of building the matrix $B_k$ explicitly,  the action of inverse Hessian matrix applied to the gradient vector, is computed to obtain the descent direction
$d_k:=-B_k\nabla  f^k$ based on the two-loop recursion \cite[algorithm 7.4]{Nocedal_2006_NOO}. In case the norm of the gradient vanishes or no gradient stored in memory at the first iteration, the LBFGS algorithm switches to steepest descent method by using the negative gradient as the descent direction.

\begin{lstlisting}
/*< calculate search direction (two-loop recursion) >*/
void lbfgs_descent(int n, float *g, float *d, float **sk, float **yk, float *q, float *rho, float *alpha, opt_t *opt)
{
  int i, j;
  float tmp0, tmp1, gamma, beta;

  //safeguard a descent direction from negative gradient, d=-g
  tmp0=opt->kpair>0?l2norm(n, opt->sk[opt->kpair-1]):0;
  tmp1=opt->kpair>0?l2norm(n, opt->yk[opt->kpair-1]):0;
  if(!( tmp0>0. && tmp1>0.)){
    flipsign(n, g, d); //descent direction= -gradient
    return;
  }

  //store the gradient in vector q
  memcpy(q, g, n*sizeof(float));
    
  //first loop
  for(i=opt->kpair-1; i>=0; i--){
    // calculate rho
    tmp0=dotprod(n, yk[i], sk[i]);
    tmp1=dotprod(n, sk[i], q);
    rho[i]=1./tmp0;
    alpha[i]=rho[i]*tmp1;
    for(j=0; j<n; j++) q[j] -= alpha[i]*yk[i][j];
  }

  tmp0 = 1./rho[opt->kpair-1];
  tmp1 = dotprod(n, yk[opt->kpair-1], yk[opt->kpair-1]);
  gamma = tmp0/tmp1;//initial Hessian = gamma* I
  for(j=0; j<n; j++) d[j]=gamma*q[j];

  //second loop
  for(i=0; i<opt->kpair; i++){
    tmp0=dotprod(n, yk[i], d);
    beta=rho[i]*tmp0;
    tmp1=alpha[i]-beta;
    for(j=0; j<n; j++) d[j] += tmp1*sk[i][j];
  }

  for(j=0; j<n; j++) d[j]=-d[j];//descent direction = - H^(-1)*g
}
\end{lstlisting}
In the above, the memory length $\ell$ is specified by the parameter \verb|opt->kpair|; \verb|l2norm()| and \verb|dotprod()| calculate the $L_2$ norm of a vector and the dot product between two real-valued vectors, respectively.

\subsection{Bisection-based backtracking line search}

Since the LBFGS algorithm is not scale invariant, we have to determine a proper stepsize $\alpha_k$ by line search method, in order to satisfy the following two Wolfe conditions:
\begin{subequations}
  \begin{align}
     f(m^k + \alpha_k d_k)\leq & f(m^k) + c_1 \alpha_k(\nabla  f^k)^\mathrm{T} d_k,\label{eq:wolfe1}\\
    \nabla  f(m^k + \alpha_k d_k)^\mathrm{T} d_k \geq& c_2 (\nabla f^k)^\mathrm{T} d_k,\label{eq:wolfe2}
  \end{align}
\end{subequations}
where the two constants are taken to be $c_1=10^{-4}$ and $c_2=0.9$. The condition \eqref{eq:wolfe1} ensures that there is sufficient decrease of the misfit, while the condition \eqref{eq:wolfe2} enforces a consistency in the curvature to prevent extremely small stepsize.

In practice, we use a bisection strategy to repeatedly shrink the range of a proper stepsize $\alpha$ between a lower bound $\alpha_1$ and an upper bound $\alpha_2$, which varies from one iteration to the next. At the outset of line search, we initialize $\alpha_1=0$, $\alpha_2=+\infty$ and $\alpha=1$. In each iteration, we check the following things in order: If the condition \eqref{eq:wolfe1} is violated, the upper bound of $\alpha$ is set to be $\alpha_2=\alpha$ while we take $\alpha=(\alpha_1+\alpha_2)/2$; otherwise, if the condition \eqref{eq:wolfe2} is violated, the lower bound of $\alpha$ is updated to be $\alpha_1=\alpha$ and a new $\alpha$ is picked up:
\begin{equation}
  \alpha=\begin{cases}
  2\alpha, & \mbox{if} \; \alpha_2=+\infty\\
  \frac{1}{2}(\alpha_1+\alpha_2), & \mbox{otherwise}.
  \end{cases}
\end{equation}
The checking is performed iteratively until the total number of line search \verb|opt->nls| is reached.  In practice, it is also important to apply bound constraints (which are determined based on a priori knowledge of physics) in nonlinear optimization. The following code snippet serves as a precise implementation of this bisection line search strategy with bound constraints.

\begin{lstlisting}
void line_search(int n,//dimension of x
		 float *x,//input vector x
		 float *g,//gradient of misfit function
		 float *d,//descent direction
		 opt_fg fg,//subroutine to evaluate function and gradient
		 opt_t *opt)//pointer of l-BFGS 
/*< bisection line search  based on Wolfe condition >*/
{
  int j;
  float gxd, c1_gxd, c2_gxd, fcost, fxx, alpha1, alpha2;
  float *xk;
  static float infinity = 1e10;

  opt->alpha = 1.;
  alpha1 = 0;
  alpha2 = infinity;
  
  xk = alloc1float(n);//allocate memory for current x
  memcpy(xk, x, n*sizeof(float));//store x at k-th iteration
  //m3=slope of the function of alpha along direction d
  gxd = dotprod(n, g, d);//<G[f(x)]|d>
  c1_gxd = opt->c1*gxd;//c1*<G[f(x)]|d>
  c2_gxd = opt->c2*gxd;//c2*<G[f(x)]|d>
  for(opt->ils=0; opt->ils<opt->nls; opt->ils++){
    for(j=0; j<n; j++) x[j] = xk[j] + opt->alpha*d[j];//update x

    //clip x by lower+upper bounds (l-BFGS-B, bounded l-BFGS)
    if(opt->bound==1) boundx(x, n, opt->xmin, opt->xmax);
    fcost = fg(x, g);//function and gradient evaluation
    opt->igrad++;//update counter of function + gradient evaluation 

    //m3=the slope of the function of alpha along search d
    gxd = dotprod(n, g, d);//<G[f(x+alp*d)]|d>
    fxx = opt->fk + opt->alpha * c1_gxd;

    //check Wolfe condition for current step length
    //condition 1: f(x + alp*d) <= f(x) + m1*alpha*<G[f(x)]|d>
    //condition 2: <G[f(x+alp*d)]|d>  >= m2*<G[f(x)]|d>
    if(fcost > fxx){
      if(opt->verb) printf("Wolfe condition 1 fails: insufficient misfit decrease!\n");
      alpha2 = opt->alpha;
      opt->alpha = 0.5*(alpha1+alpha2);//shrink search interval
    }else if(gxd < c2_gxd){
      if(opt->verb) printf("Wolfe condition 2 fails: stepsize is too small!\n");
      alpha1 = opt->alpha;
      if(alpha2<infinity)
         opt->alpha = 0.5*(alpha1 + alpha2);//shrink search interval
      else
         opt->alpha *= 2.0;//extend search interval
    }else{//conditions satisfied, terminate line search
      break;
    }

    if(opt->verb){
      printf("#line search %d, alp1=%f alp2=%f alp=%f\n", opt->ils, alpha1, alpha2, opt->alpha);
      printf("-----------------------------------\n");
    }
  }

  if(fcost <= opt->fk) {
    opt->ls_fail = 0;
    opt->fk = fcost;//fcost not increased, accept it
  }else{
    opt->ls_fail = 1; //line search fails, exit
  }
  
  free1float(xk);
}
\end{lstlisting}

\subsection{Function and gradient evaluation with reverse communication}

To build the multigrid module in section~\ref{sec:multigrid}, we have tacitly introduced a new programming paradigm called \emph{reverse communication}. 
It creates a reliable mechanism, allowing the program easily jumping out and getting back to the same breakpoint without loss any information after the completion of the previous computational segment. This turns out to be critical to design a clean nonlinear inversion scheme, which separates the functionality of quasi-Newton optimization with function and gradient evaluation at each iteration. Indeed, reverse communication has been widely used in Fortran programming  \cite{markus2012modern}. The programmers usually introduce a flag parameter in Fortran to achieve reverse communication: the code calls the same subroutine but taking different values for the flag in order jump out of the computational segment after a while, or jump into it to continue an uncompleted procedure.

Different from Fortran convention, we have decided to separate them in our C code. The module for function and gradient evaluation is divided into three different subroutines:
\begin{enumerate}
\item Once the inversion starts, we create local copies of the pointers to access the information of acquisition geometry stored in pointer \verb|acq|, electromagnetic fields indexed by pointer \verb|emf| and inversion parameters linked with the pointer \verb|fwi|, by calling the routine
\begin{verbatim}
void fg_fwi_init(acq_t *acq_, emf_t *emf_, fwi_t *fwi_);
\end{verbatim}
which mimics the constructor in C++ class.

\item The nonlinear inversion updates the model parameters iteratively. During each iteration, we resort to the routine
\begin{verbatim}
float fg_fwi(float *x, float *g)
\end{verbatim}
for function and gradient evaluation, where the input model vector and output gradient vector are stored in \verb|x| and \verb|g|, with a returned value being the computed misfit. We highlight that there are many places that require accessing to the pointers \verb|acq|, \verb|emf| and \verb|fwi|, which can be done internally thanks to the local copies of these pointers initialized by \verb|fg_init(...)|. This makes the argument list of \verb|fg_fwi| rather compact so that it would be called frequently among different iterations.

\item At the end of the inversion, the variables used in this module will be deallocated via \verb|fg_close()|, which plays the role of a destructor as in C++ class.
\end{enumerate}

\begin{figure}[!htbp]
  \centering
  \includegraphics[width=0.9\linewidth]{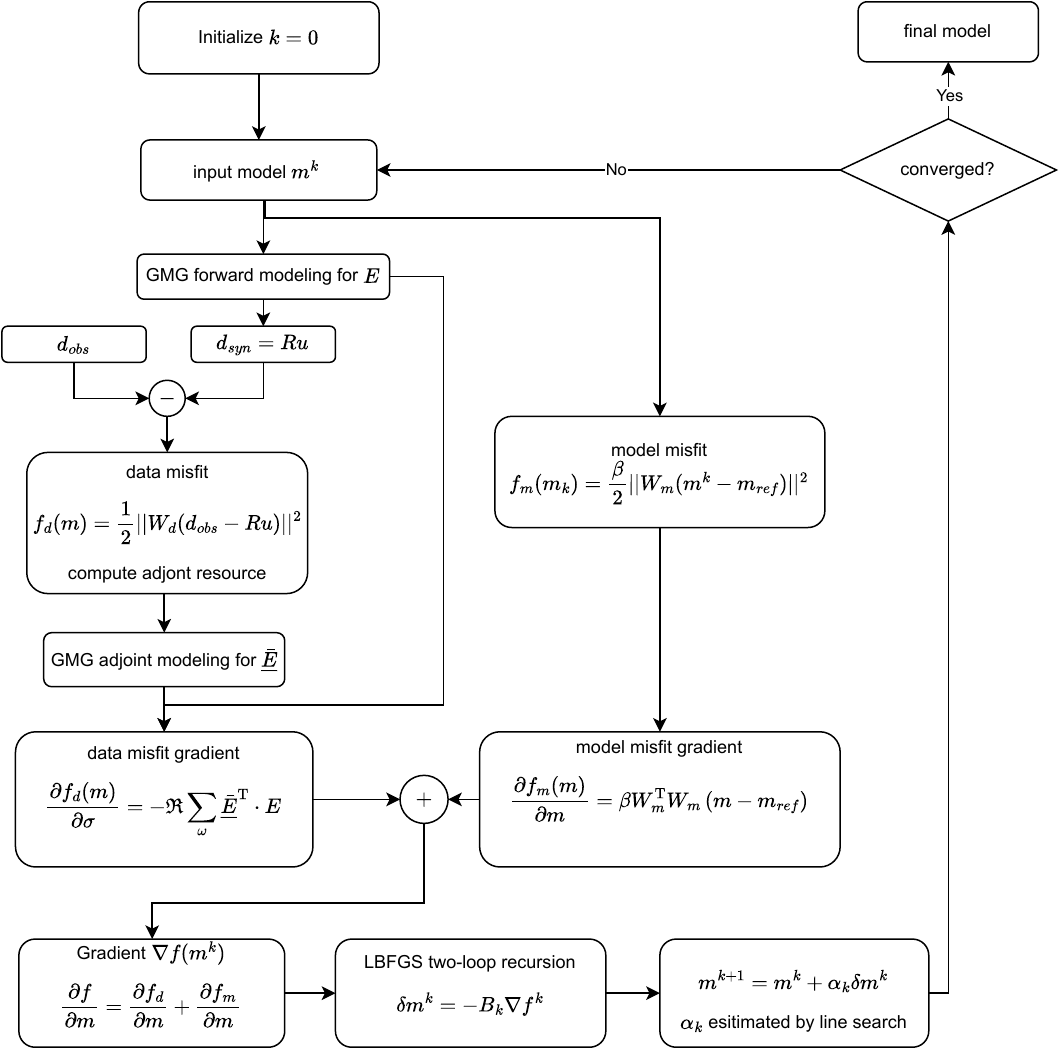}
  \caption{3D CSEM inversion workflow using frequency-domain GMG modelling}\label{fig:workflow}
\end{figure}

The reverse communication allows us to greatly simplify the preconditioned LBFGS algorithm for nonlinear optimization, as illustrated  in the following:
\begin{lstlisting}
  for(opt->iter=0; opt->iter<opt->niter; opt->iter++){
    ...    
    memcpy(opt->q, opt->g, fwi->n*sizeof(float));
    if(opt->iter==0){/* first iteration, no stored gradient */
      if(opt->preco) precondition(emf, fwi, opt->q);
      flipsign(fwi->n, opt->q, opt->d);//descent direction=-gradient
    }else{
      lbfgs_update(fwi->n, opt->x, opt->g, opt->sk, opt->yk, opt);
      opt->rho = alloc1float(opt->kpair);
      opt->alp = alloc1float(opt->kpair);
      if(opt->preco) {
        lbfgs_descent1(fwi->n, opt->g, opt->q, opt->rho, opt->alp, opt->sk, opt->yk, opt);
        precondition(emf, fwi, opt->q);
        if(opt->loop1) lbfgs_descent2(fwi->n, opt->g, opt->q, opt->rho, opt->alp, opt->sk, opt->yk, opt);
      }else
        lbfgs_descent(fwi->n, opt->g, opt->d, opt->sk, opt->yk, opt->q, opt->rho, opt->alp, opt);
      free1float(opt->rho);
      free1float(opt->alp);
    } 
    lbfgs_save(fwi->n, opt->x, opt->g, opt->sk, opt->yk, opt);
    line_search(fwi->n, opt->x, opt->g, opt->d, fg_fwi, opt);
    ...
  }
\end{lstlisting}
Here, the routine \verb|lbfgs_update(...)| updates  the storage to keep only the latest \verb|opt->kpair| pairs of $s_k$ and $y_k$. The estimation of the descent direction using two-loop recursion LBFGS algorithm by \verb|lbfgs_descent()| can be split into two subroutines \verb|lbfgs_descent1()| and \verb|lbfgs_descent2()|. In between, a preconditioner can be applied to achieve a preconditioned LBFGS implementation. 
The function name \verb|fg_fwi| has been used as one of the arguments in \verb|line_search| stage, thanks to a function pointer defined in the type of
\begin{verbatim}
typedef float (*opt_fg)(float*, float*); 
\end{verbatim}
Putting all these techniques together, Figure~\ref{fig:workflow} schematically illustrates the complete workflow of 3D CSEM inversion using GMG modelling engine.

\section{Numerical examples}

\subsection{Forward modelling}

\begin{figure}[!htbp]
  \centering
  \includegraphics[width=0.6\linewidth]{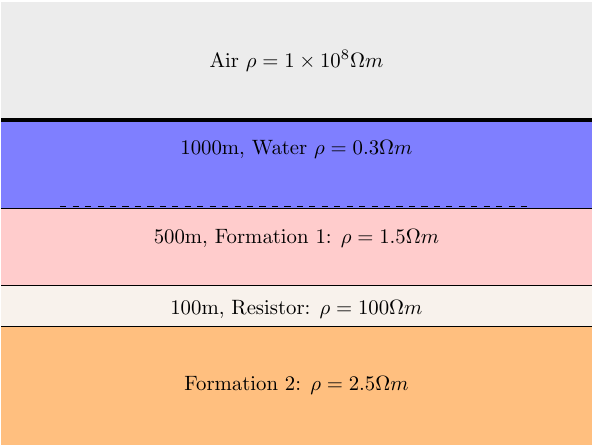}
  \caption{The 1D layered resistivity model}\label{fig:model1d}
\end{figure}

We now perform a comparative study on the numerical modelling of \verb|libEMMI_MGFD| in a 1D layered resistivity model as shown in Figure~\ref{fig:model1d}. In this simulation, we deploy a x-directed transmitter at the water depth of 950 m, while a line of receivers are placed at the sea floor (50 m below the source).  This allows us to check the accuracy of our modelling using 1D semi-analytic solution thanks to the \verb|empymod| program \cite{werthmuller2017open}. Because \verb|libEMMI_MGFD| performs  the same multigrid modelling as \verb|emg3d| code \cite{werthmuller2019emg3d} does, it is also instructive to make a cross-validation among the three. In order to make a fair comparison, we input the resistivity model of exactly the same size (\verb|nx=ny=200|, \verb|nz=100|) and grid spacing (\verb|dx=dy=100|, \verb|dz=40|) for \verb|emg3d| and \verb|libEMMI_MGFD|. The code will make an extended model of proper size for efficient GMG modelling while avoiding the boundary effect.

\begin{figure}
  \centering
  \subfloat[$E_x$]{\includegraphics[width=0.49\linewidth]{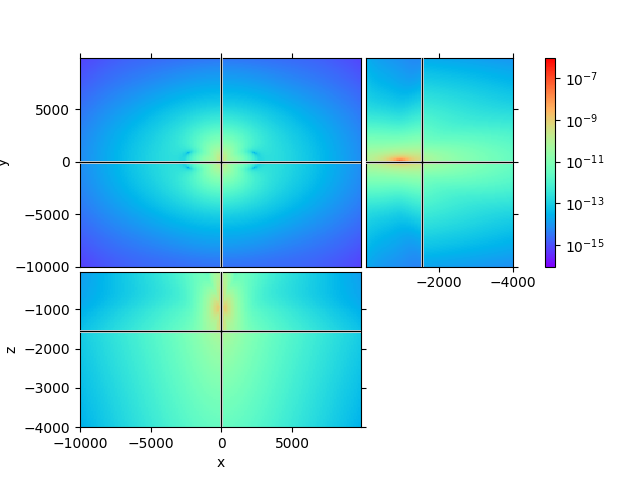}}
  \subfloat[$H_x$]{\includegraphics[width=0.49\linewidth]{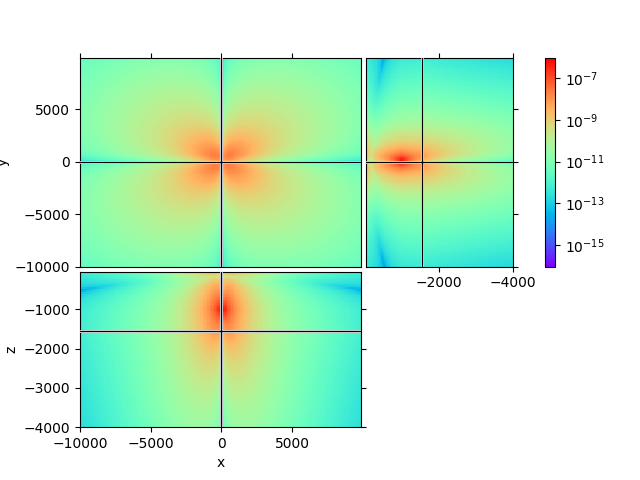}}\\
  \subfloat[$E_y$]{\includegraphics[width=0.49\linewidth]{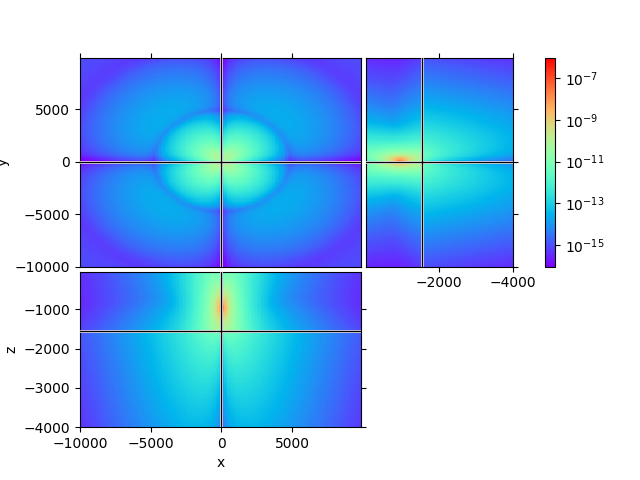}}
  \subfloat[$H_y$]{\includegraphics[width=0.49\linewidth]{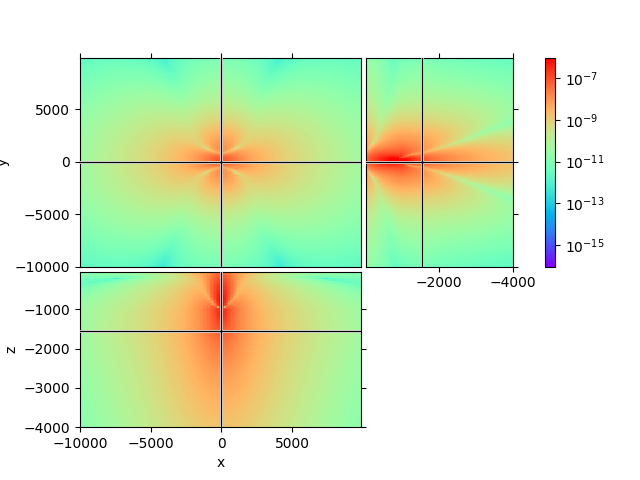}}\\
  \subfloat[$E_z$]{\includegraphics[width=0.49\linewidth]{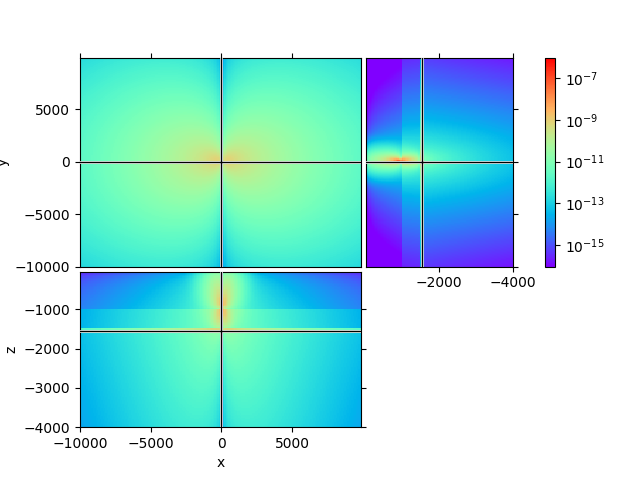}}
  \subfloat[$H_z$]{\includegraphics[width=0.49\linewidth]{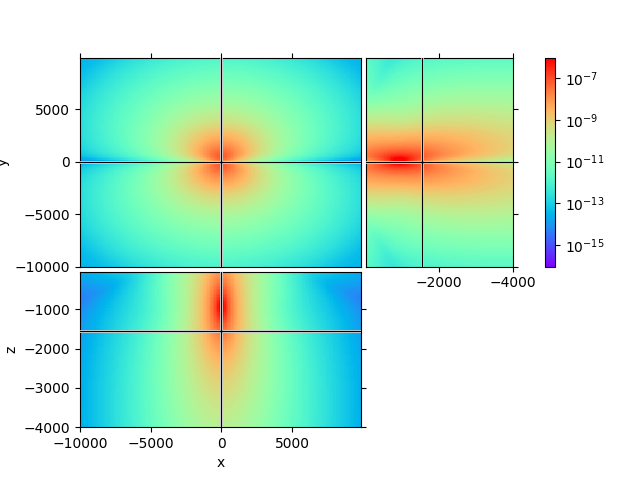}}\\
  \caption{The amplitude of the EM fields modelled in the 1D layered model}\label{fig:fields}
\end{figure}

\begin{figure}[!htbp]
  \centering
  \includegraphics[width=0.9\linewidth]{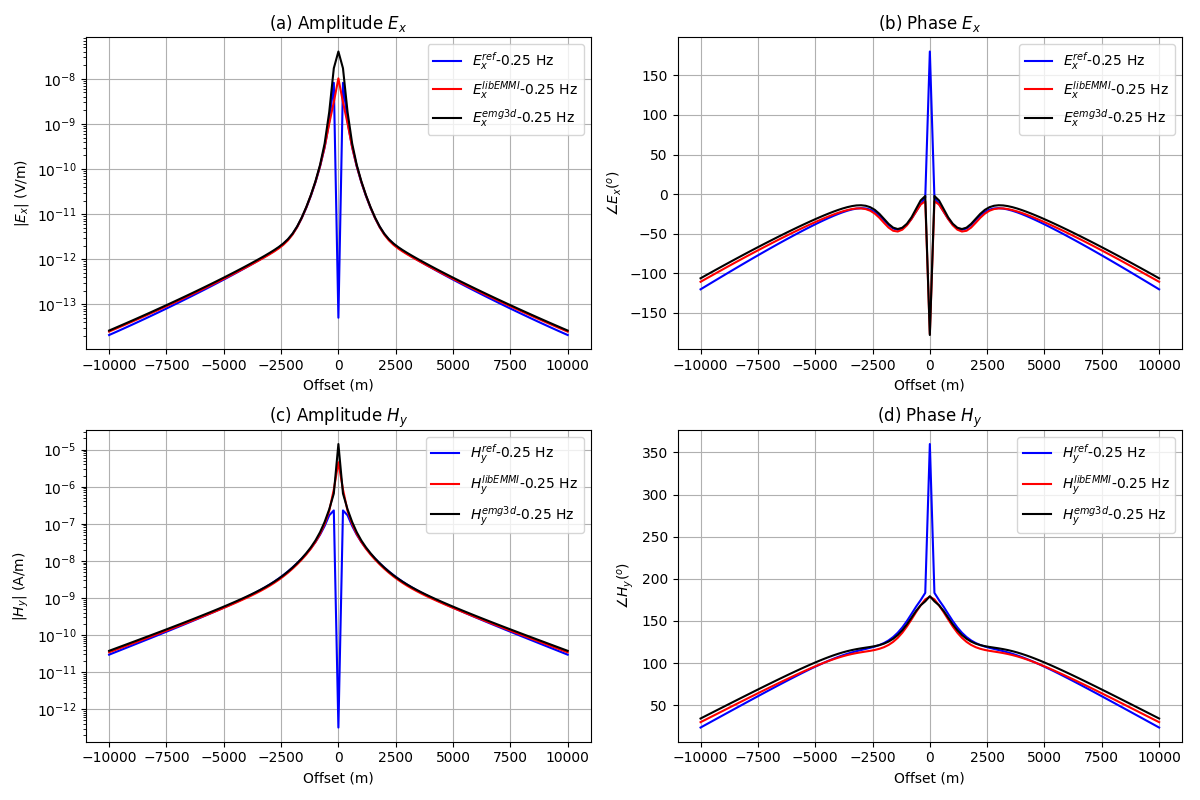}
  \caption{Comparison between the semi-analytic solution with the numerical solution in a 1D layered model}\label{fig:data}
\end{figure}

Figure~\ref{fig:fields} shows all the electromagnetic field components after the modelling using \verb|libEMMI_MGFD|. To quantitatively check the amplitude and phase of our modelling, we compare the $E_x$ and $H_y$ component of the EM data recorded by the receivers among \verb|libEMMI_MGFD|, \verb|emg3d| and \verb|empymod| in Figure~\ref{fig:data}. We have seen a very good agreement among these three in terms of both amplitude and phase. The phases in Figure~\ref{fig:data}b and \ref{fig:data}d seem to indicate that compared to the solution from \verb|emg3d|, the solution from \verb|libEMMI_MGFD| is slightly closer to the semi-analytic solution obtained from \verb|empymod|.

Let us also check the computational time and memory consumption between \verb|emg3d| and \verb|libEMMI_MGFD|. To complete the above simulation, both \verb|emg3d| and \verb|libEMMI_MGFD|  extend the input model to a size of $256\times 256\times 160$ for multigrid solving using semi-coarsening and line relaxation. In terms of CPU time, \verb|emg3d| converges within 16.2 minutes after 13 F cycles, while \verb|libEMMI_MGFD| converges with 14.7 minutes after 30 F cycles.
Taking into account the actual number of floating point operations, the computation by \verb|libEMMI_MGFD| coded in C is considered much faster than that of \verb|emg3d| coded in python. This verifies that directly programming in C can still be beneficial in terms of efficiency, while the python programming adding JIT (just-in-time) decorator can create a code with significant speedup so that the performance of the compiled executable is very close to a C/Fortran program of the same functionality.  We notice that \verb|libEMMI_MGFD| consumes 3.5 GB memory while \verb|emg3d| only requires 2.8 GB. This may be caused by the fact that \verb|libEMMI_MGFD| uses ILU factorization without exploiting the benefit of symmetry of Cholesky decomposition as done in \verb|emg3d|, besides the additional memory allocation to facilitate inversion.

\begin{figure}[!htbp]
  \centering
  \includegraphics[width=0.9\linewidth]{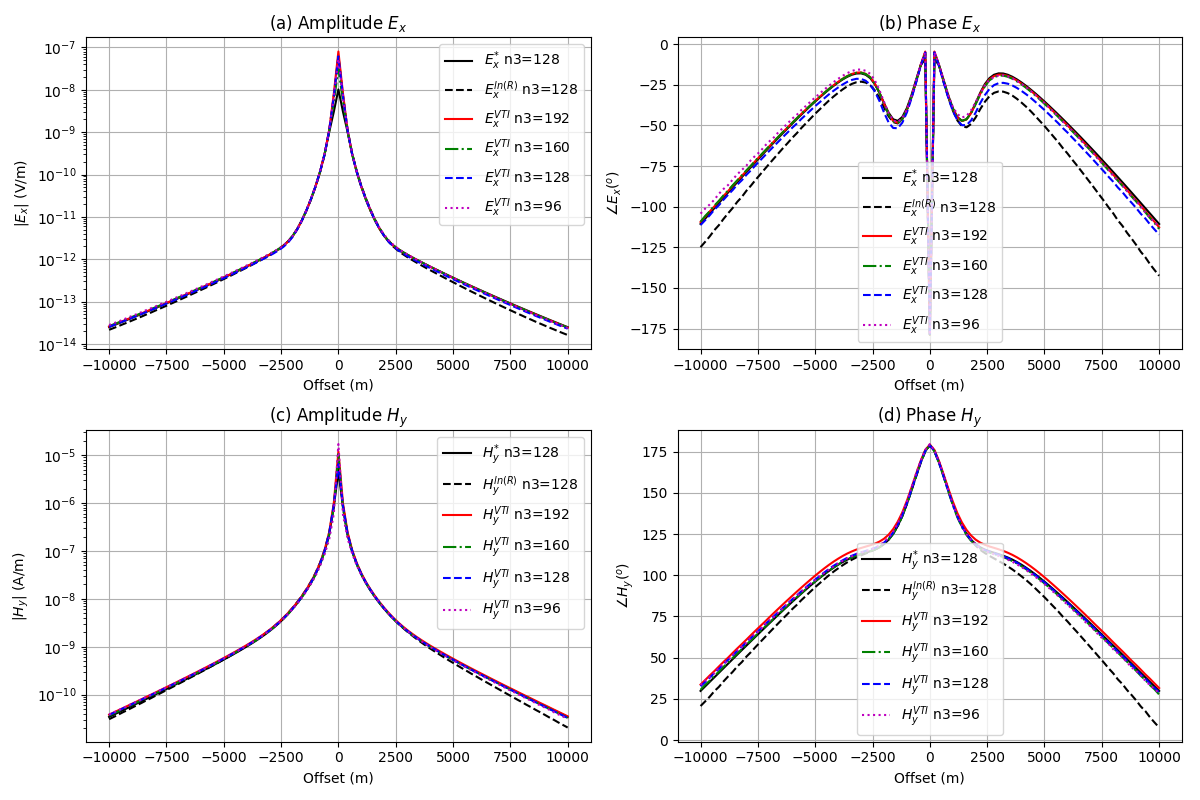}
  \caption{The simulated EM data using effective medium by gridding the 1D model with different number of intervals in the z direction: E/H superscript $^*$ corresponds to the solution without medium homogenization; E/H with superscript $VTI$ are modelled using effective medium by VTI anisotropic averaging; E/H with superscript $\ln(R)$ are modelled using homogenized medium by averaging over logarithm of the conductivity as in \cite{plessix2007approach}.}\label{fig:averaging}
\end{figure}

We now compare the solution of \verb|libEMMI_MGFD| using different gridding strategies by activating automatic gridding using our VTI anisotropic averaging. Figure~\ref{fig:averaging} shows that the the modelled EM fields by VTI anisotropic averaging after regridding the modelling into different number of intervals along z direction still matches the numeric solution without any averaging. However, the simulated EM field via averaging over logarithm of the conductivity proposed in \cite{plessix2007approach} seems to drift away quite a lot, which is very visible in the phase diagram of both electric and magnetic fields.

\begin{figure}[!htbp]
  \centering
  \includegraphics[width=0.9\linewidth]{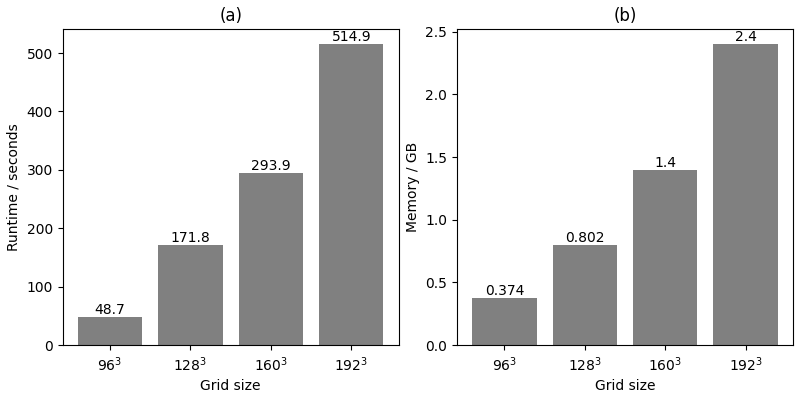}
  \caption{The runtime and memory for modelling in 1D layered model}\label{fig:runtime}
\end{figure}

After confirming the consistency of the modelling using homogenized effective medium, we are now interested in making an assessment of the computationally efficiency gained from regridding the model into a reduced size. Figure~\ref{fig:runtime}a shows that the runtime for modelling  grows quickly with the increasing of the mesh size. Meanwhile, the consumption of the computer memory exhibits a similar increasing trend as shown in Figure~\ref{fig:runtime}b. These numerical tests suggest that working with the effective medium by VTI anisotropic averaging is advantageous in terms of computational efficiency and memory consumption, as long as the model is regridded properly without loss of too much details.

\subsection{Resistivity imaging}

\begin{figure}[!htbp]
  \centering
  \subfloat[]{\includegraphics[width=0.49\linewidth]{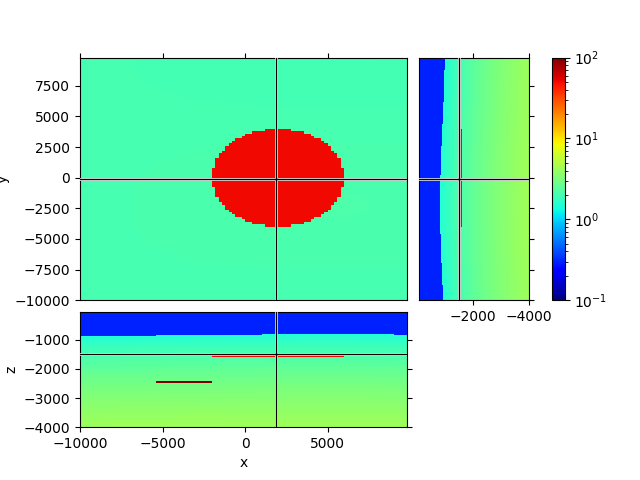}}
  \subfloat[]{\includegraphics[width=0.49\linewidth]{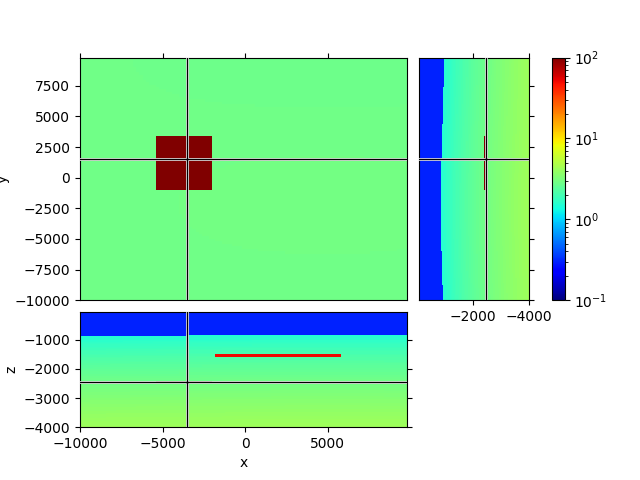}}\\
  \subfloat[]{\includegraphics[width=0.49\linewidth]{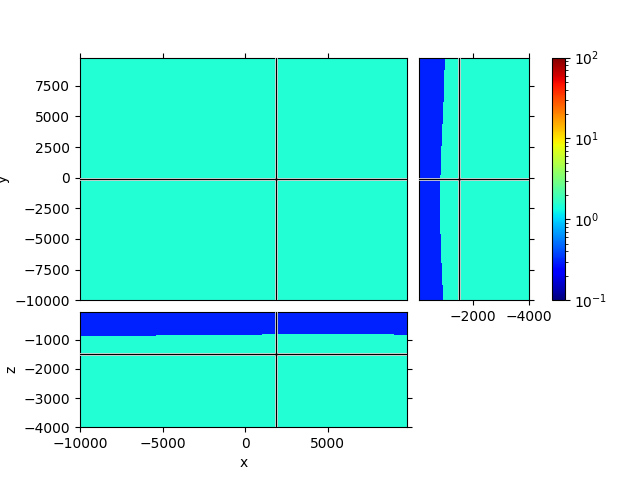}}
  \subfloat[]{\includegraphics[width=0.49\linewidth]{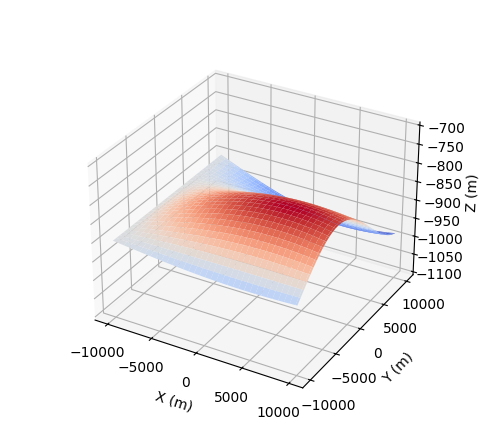}}
  \caption{The true resistivity model contains two resistors in the shape of disk (a) and square (b), while the initial model takes a homogeneous background of 1.5 $\Omega\cdot m$ below the sea water of 0.3 $\Omega\cdot m$. The sea water and the formation are separated by the bathymetry in (d). }\label{fig:model}
\end{figure}

Using the GMG modelling engine, we perform a synthetic 3D marine CSEM inversion for resistivity imaging. The true model includes two resistors in the shape of a disk (Figure~\ref{fig:model}a) and a square (Figure~\ref{fig:model}b) sitting at different depth. The model expands horizontally over 20 km in both x and y directions. We start from a crude initial model with a homogeneous background of 1.5 $\Omega\cdot m$ below the sea water which is of 0.3 $\Omega\cdot m$. The sea water and the sediment of the Earth are separated by a seafloor with depth variations as depicted in Figure~\ref{fig:model}d.

\begin{figure}[!htbp]
  \centering
  \includegraphics[width=0.75\linewidth]{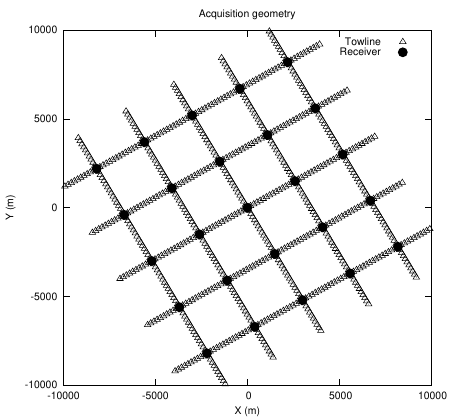}
  \caption{Survey layout sheet}\label{fig:survey}
\end{figure}

To mimic a real acquisition, there are 10 towlines of 16 km  deployed: half of the towlines (up to an azimuth angle of 30 degree compared to computing coordinate) are orthogonal to the other ones.  In total, 25 sources are places at the cross of the towlines at the seabed. Along each towline, the receivers are distributed with equal spacing of 200 m and at least 50 m above the seabed.
Note that the the number of sources are much larger than the number of receivers in real CSEM data acquisition. In practice the reciprocity is often applied to form a collection of common receiver gathers as the virtual sources, in order to drastically reduce the computational expense. Our experiment mimics the configuration of imaging the real CSEM data acquired with a source-receiver geometry depicted in Figure~\ref{fig:survey} after switching sources and receivers.

We generate the observed data using the true resistivity model for three frequencies: 0.25 Hz, 1.0 Hz and 2.75 Hz. When simulating multiple frequencies, we observed that the GMG solver converges much faster at higher frequencies. This makes sense because the decay rate of the EM fields becomes larger at the higher frequency band.

\begin{figure}[!htbp]
  \centering
  \includegraphics[width=\linewidth]{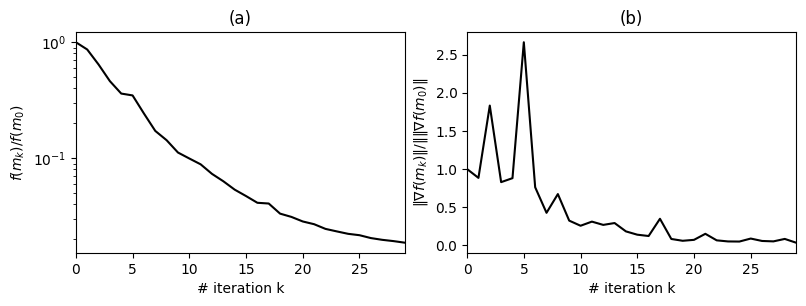}
  \caption{The evolution of  (a) the normalized misfit and (b) the norm of the gradient in CSEM inversion}\label{fig:misfit}
\end{figure}

Using the initial model in Figure~\ref{fig:model}c, we run the 3D VTI inversion using LBFGS algorithm for 30 iterations. Since the EM data modelled using a total field approach are not precise at the near offset, the near field within 5.5 skin depth for different frequencies are muted during the inversion.
A depth weighting \cite{Plessix_2008_RIC} has been used to precondition the LBFGS optimization. Because the energy of the diffusive EM fields decays quickly along the depth, such a weighting helps to boost the weak model update in the depth.  The data misfit has been reduced down to less than 2\% of the misfit at the 1st iteration, while the norm of the gradient is also greatly reduced, as shown in Figure~\ref{fig:misfit}. The root-mean-square error (RMSE) has been reduced from 8.9 in the first iteration to 1.2 after 30 iterations.

\begin{figure}[!htbp]
  \centering
  \subfloat[$\rho_v$]{\includegraphics[width=0.49\linewidth]{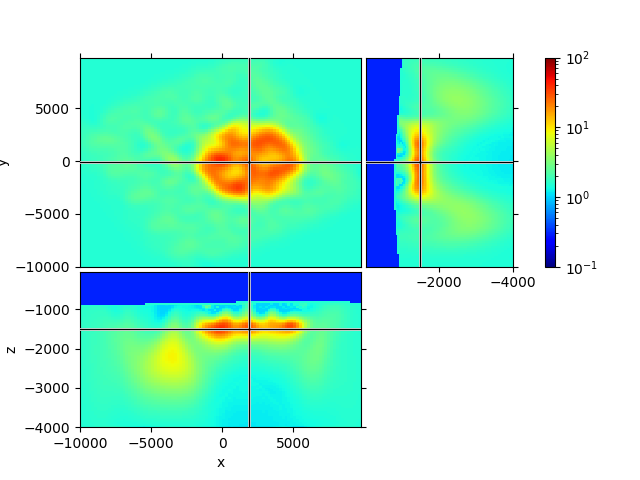}}
  \subfloat[$\rho_h$]{\includegraphics[width=0.49\linewidth]{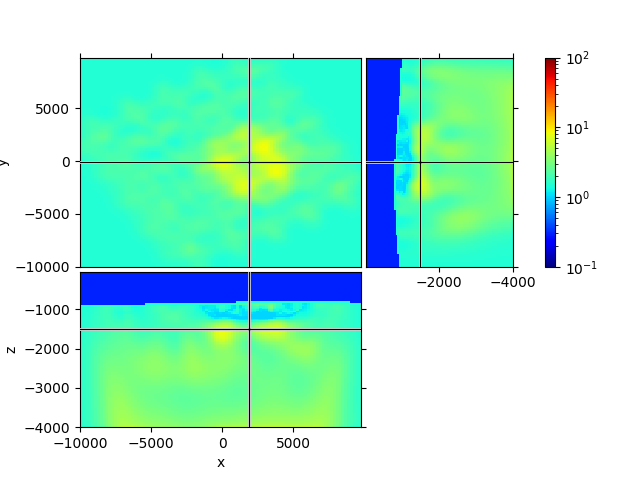}}
  \caption{Inverted resistivity models}\label{fig:inv}
\end{figure}

Figure~\ref{fig:inv} shows the reconstructed vertical resistivity and horizontal resistivity. Figure~\ref{fig:inv}a reveals that the resistor in the shallow part has been much better resolved than the deep resistor. Meanwhile, the horizontal resistivity in Figure~\ref{fig:inv}b seems to show significant amount of update in the background, despite that the resistive anomalies are reconstructed poorly. It is worthy noting that the reconstruction of the resistors in horizontal directions is much better than the vertical direction, in which a strong blurring effect with the increase of the depth in the retrieved resistive anomalies  can be observed.

\begin{figure}[!htbp]
  \centering
  \includegraphics[width=\linewidth]{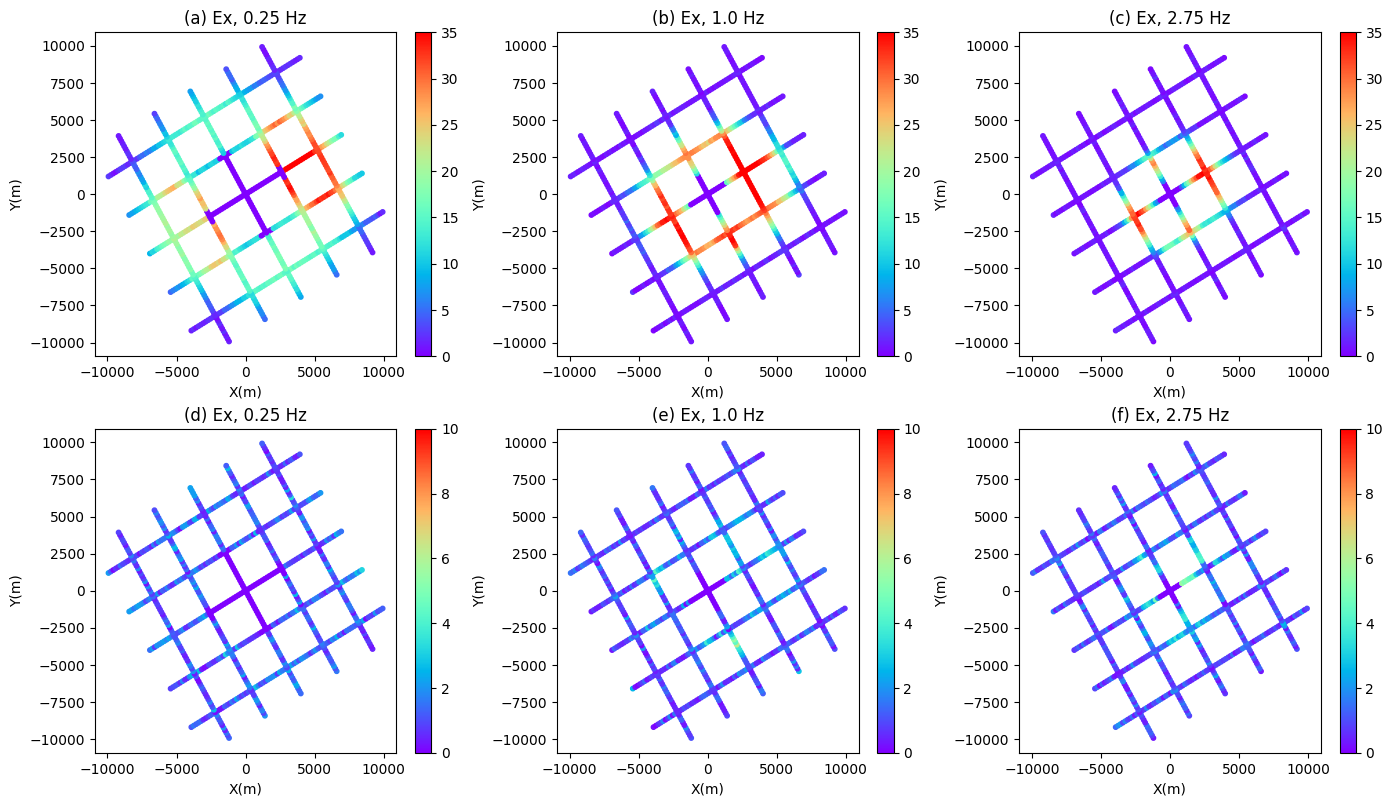}
  \caption{Comparison of RMSE misfit for a source at the center of the domain before (a,b,c) and after (d,e,f) 3D CSEM inversion over three frequencies}\label{fig:rmse}
\end{figure}

To further check the quality of the data matching, we plotted the RMSE between the observed and synthetic data using the initial model and inverted model in Figure~\ref{fig:rmse} for all three frequencies corresponding to a source at the center of the model. Before CSEM inversion, the RMSE misfit for 0.25 Hz, 1 Hz and 2.75 Hz are large, as can be seen from the top panels of Figure~\ref{fig:rmse}. They have been significantly decreased after 3D inversion, as shown in the bottom panels of Figure~\ref{fig:rmse}.

\section{Discussions}

In the current work, we focus on a pure frequency domain solution of diffusive Maxwell equation using multigrid iterative solver.
This solver must be performed frequency by frequency in a sequential manner. The benefit of a time domain solver is that it computes all frequencies in one go, but the efficiency of such a scheme strongly depends on the Courant-Friedrichs-Lewy (CFL) stability condition, which limits the size of grid spacing and temporal step size. For 3D resistivity tomography targeting to the oil and gas industry, the CSEM inversion starts with a conductive background model, in which a time domain solver is advantageous. With the evolution of inversion, the model parameters exhibit some resistive anomalies, making the computational efficiency of frequency-domain GMG solver potentially superior to time domain solver. We aim to combine both time and frequency domain approaches into a hybrid scheme for more efficient inversion, by a reorganization of \verb|libEMMI| and \verb|libEMMI_MGFD| in the future.

Motivated by the reduction of memory consumption, BICGSTAB has not been included using GMG as a preconditioner, although it has been recommended by \cite{mulder2006multigrid}. We realize that the modelling accuracy of current GMG solver needs to be improved further. In a 1D resistivity model, we have observed that the modelling accuracy by current GMG engine is much lower compared to high order fictitious wave domain time-stepping modelling. The accuracy of the electrical field in the current modelling engine is limited to 2nd order accuracy, while the magnetic field is only 1st order. This may be improved by adapting GMG into a higher order scheme in the near future.

Two different sets of grid are sometimes used for 3D CSEM simulation and inversion \cite{Plessix_2008_RIC}:  The inversion updates the resistivity parameters on a uniform grid where the input parameters are given, while  the modelling runs over a nonuniform grid created based on the homogenized effective medium. Since the modelling grid is source dependent and difficult to be consistent, the gradient is therefore built using the extracted EM fields at the inversion grid by interpolation in the modelling grid.
Despite the ease of such a configuration in \verb|libEMMI_MGFD| and the efficacy of simulation over nonuniform grid,
our numerical experience indicates that a uniform gridding for both modelling and inversion helps avoid some inconsistent model update from different sources and the instability in decreasing the misfit induced by inconsistent gridding, though the computational cost is increased correspondingly. We have also tested \verb|libEMMI_MGFD| by simultaneous inversion of multicomponent CSEM data. Dramatic reduction of the data misfit  seems to be more difficult for multicomponent than single component inversion. The inversion produces a resistivity model less satisfactory than the single component inversion presented above. It is expected to be improved by developing a higher order GMG modeling engine.

The Gauss-Newton method (a type of Newton-CG method) is another popular choice in CSEM inversion \cite{li2010three,zaslavsky2013large,haber2014parallel,peng20153d,amaya20163d,mittet2024gauss}. The Gauss-Newton algorithm requires  two levels of nested loops: an inner loop using linear conjugate gradient method to compute the descent direction by iterative solution of the normal equation according to a symmetric positive Hessian matrix, and an outer loop to do nonlinear update for the model properties. In \verb|libEMMI_MGFD|, the subroutine
\begin{verbatim}
void cg_solve(int n, //dimension of x
	      float *x, //input vector x
	      float *g, //gradient of misfit function
	      float *d, //descent direction
	      opt_Hv Hv, //subroutine to evaluation function and gradient
	      opt_t *opt) //pointer of optimization 
\end{verbatim}
has been designed to accomplish the inner linear inversion through CG. As long as the Hessian vector product can be computed in a matrix free manner via a dedicated routine of the following form
\begin{verbatim}
void Hv_fwi(float *x, float *v, float *Hv){ ...}
\end{verbatim}
one can perform Gauss-Newton inversion by feeding \verb|cg_solve| with an input parameter wich is the name of the routine of the type
\begin{verbatim}
typedef void (*opt_Hv)(float*, float *, float*);
\end{verbatim}
This Newton-CG method has been investigated in the seismic full waveform inversion settings \cite{Yang_2018_TRN}. The study in \cite{Yang_2018_TRN} shows that the global computational cost of Gauss-Newton method is generally higher than quasi-Newton LBFGS algorithm, despite the improved convergence according to the number of outer loops.
 The quasi-Newton LBFGS algorithm is therefore chosen as the default optimization method in \verb|libEMMI_MGFD| thanks to its efficiency.

\section{Conclusion}

We have developed a CSEM modelling and inversion software \verb|libEMMI_MGFD| using a frequency-domain multigrid solver. The multigrid modelling, constructed by V/F cycles involving Gauss-Seidel smoothing, restriction and prolongation phases, has been modularized for compactness and efficiency.  The default nonlinear optimization method is LBFGS algorithm assisted with bisection-based line search. A reverse communication mechanism gives great flexibility for function and gradient evaluation. The code is programmed in C and parallelized using MPI to deal with large scale applications for resistivity tomography. Numerical examples are presented to demonstrate the forward simulation and inverse imaging capabilities of \verb|libEMMI_MGFD|. We envision some necessary modifications to process the field data for industrial applications in the future.

\section*{Acknowledgements}

This research is financially supported by the Fundamental Research Funds for the Central Universities (Grant No. HIT.OCEF.2024025) and National Natural Science Foundation of China (Grant No. 42274156). Pengliang Yang is indebted to Dieter Werthm\"{u}ller for responsive feedback in last two years to develop the geometrical multigrid code using C programming language. Without his help and the Python code \verb|emg3d|, it is not possible to complete \verb|libEMMI_MGFD| to share with the community.

\section*{Data Availability}
No data has been used for this work.

\section*{CRediT authorship contribution statement}

\textbf{Pengliang Yang}: Conceptualization, Formal analysis, Investigation, Methodology, Software, Validation, Writing, Supervision, Project administration, Funding acquisition.
\textbf{An Ping}: Methodology, Software, Validation, Writing – review \& editing.

\section*{Declaration of Competing Interest}
We declare that we have no financial and personal relationships with other people or organizations that can inappropriately influence our work.

\bibliographystyle{elsarticle-num}
\newcommand{\SortNoop}[1]{}

\end{document}